\documentclass[a4paper,fleqn]{cas-sc}
\pdfoutput=1
\usepackage[numbers]{natbib}
\usepackage{array}
\usepackage{subcaption}
\usepackage{hyperref}
\usepackage{orcidlink}
\usepackage{url}
\usepackage{multirow}

\usepackage{xcolor}

\begin{document}
\let\WriteBookmarks\relax
\def\floatpagepagefraction{1}
\def\textpagefraction{.001}

\shorttitle{Transverse Single-Spin Asymmetry Extraction}

\shortauthors{Pate, Arachchige, Kuruppu, and Nawarathne}  

\newcolumntype{A}{>{\raggedright\arraybackslash}m{9cm}}
\newcolumntype{L}{>{\centering\arraybackslash}m{2.5cm}}

\graphicspath{ {./imgs/} }

\title[mode = title]{Transverse Single-Spin Asymmetry Extraction, using both Binned and Unbinned Analysis, including Background Subtraction and Unfolding}
\author[1]{S.F. Pate}[orcid=0000-0001-8577-3405]
\cormark[1]
\ead{stephen.pate.physics@gmail.com}
\author[1]{H. Arachchige}[orcid=0000-0002-2251-928X]
\ead{harsha92@nmsu.edu}
\author[1]{C. Kuruppu}[orcid=0000-0003-2772-1978]
\ead{ckuruppu@nmsu.edu}
\author[1]{D. Nawarathne}[orcid=0000-0001-5395-1190]
\ead{dinupa@nmsu.edu}
\affiliation[1]{organization={Department of Physics, New Mexico State University},
            city={Las Cruces},
            postcode={88003}, 
            state={NM},
            country={USA}}

\begin{abstract}
The determination of transverse single-spin asymmetries in experiments involving polarized targets and/or beams may encounter challenges when (1) the magnitude of the polarization varies greatly with time, (2) the polarization magnitude is not the same for each spin state, (3) different integrated luminosities occur for different spin states or different target materials, and/or (4) some kinematic variables require unfolding; these are just a few examples. We present general methods of determining the asymmetry based on both binned analysis and unbinned maximum likelihood optimization, incorporating the unfolding of kinematic variables that are smeared by detector effects, and also including the possibility of background subtraction.
\end{abstract}

\begin{highlights}
\item We summarize various aspects of the determination of a transverse single-spin asymmetry, including a number of complications that may arise due to the limitations of the experimental apparatus.
\item We test our method in a case where instrumental smearing is negligible, and in two cases where the smearing is significant enough to warrant unfolding of the data.
\item In the latter cases, two different unfolding methods are used and compared.
\end{highlights}

\begin{keywords}
 data analysis \sep statistical methods \sep 
 data processing methods \sep
 data reduction methods \sep
 data unfolding methods
\end{keywords}

\maketitle

\section{Introduction}
\label{sec:intro}

In many circumstances in nuclear and particle physics, the interaction between particles includes a spin-momentum correlation which will produce an azimuthal modulation in the scattering angular distribution.  Such correlations can be studied by using a spin-polarized beam and/or target, where the direction of the polarization is normal to the beam axis.  Then the azimuthal component of the angular distribution will often take the following form,
\begin{equation}
A(\phi) \propto 1 + P A_N\sin(\phi_{\rm pol}-\phi),
\end{equation}
where $\phi$ is the azimuthal angle at which the scattered particle has been detected, $\phi_{\rm pol}$ is the azimuthal angle of the polarization direction, $P$ is the magnitude of the polarization, and $A_N$ is the {\em transverse single-spin asymmetry}\footnote{Also called the {\em analyzing power}.} caused by the physics underlying the spin-momentum correlation.  The experiment seeks to determine $A_N$ from the scattering angular distribution.  In order to eliminate a certain class of systematic effects associated with detector imperfections, it is common practice to ``flip the spin'' from time to time; for example $\phi_{\rm pol}$ might change from $+\pi/2$ (spin ``up'') to $-\pi/2$ (spin ``down'') on a given schedule.  It is generally good practice, as well, to maintain the magnitude of the polarization $P$ at a nearly constant value, but this is not always possible; in any case it is important to measure the polarization often. 
Additionally, the events of interest (called ``signal'' or ``foreground'') may be accompanied by background events with the same kinematic signature but different asymmetry.  Some model (based on simulation and/or data) must be used to remove the influence of the background in order to determine the asymmetry of the foreground correctly.  Lastly, the ability of the detector and the analysis software to reconstruct the azimuthal angle $\phi$ may be imperfect, leading to ``smearing'' or ``bin migration'' effects that need to be unfolded.

In this manuscript we will first establish both binned and unbinned methods of analysis for the determination of transverse single-spin asymmetries including the necessity of background subtraction.  Some of the material presented is ``common knowledge'' but some results are new to this publication.  Then we will include the additional steps needed in case the smearing effects are large enough to require unfolding, both for binned and unbinned analysis.

\section{Binned Estimation of the Asymmetry: without Unfolding}
\label{sec:binned_no_unfold}

We develop a general binned analysis to determine $A_N$ in the presence of a background requiring subtraction.  If we wish to include the most general case, where the polarization $P$ and integrated luminosity $L$ for each spin-state might not be the same, then we should start from the most basic expression for the foreground yield binned in the azimuthal angle $\phi$.
\begin{equation}
Y^{\pm}_F(\phi)=L^{\pm}\sigma_F[1+P^{\pm }A_F\sin(\pm\pi/2 -\phi)]
\label{eq:foreground_yield}
\end{equation}
The notation has been adjusted to distinguish foreground and background; $A_F$ corresponds to the $A_N$ mentioned in the introduction.  The subscript $F$ indicates this is the yield ($Y_F$), cross section ($\sigma_F$) and asymmetry ($A_F$) for the foreground process (the process of interest).  The ``$\pm$'' indicates whether the data was taken with the polarized target in the spin-up ($+$) or spin-down ($-$) state.  The corresponding yield from the background process (subscript $B$) would be:
\begin{equation}
Y^{\pm}_B(\phi)=L^{\pm}\sigma_B[1+P^{\pm }A_B\sin(\pm\pi/2 -\phi)]
\label{eq:background_yield}
\end{equation}
In the experiment, these foreground and background yields have occurred in the same kinematic bin and are not distinguishable from one another.  We measure a total yield (subscript $T$), which is the sum of the foreground and background.
\begin{equation}
Y^{\pm}_T(\phi)=Y^{\pm}_F(\phi)+Y^{\pm}_B(\phi)
\label{eq:total_yield}
\end{equation}
To remove the effect of the background, we must have a model for the background process determined either from the experiment itself or from a simulation based upon detailed knowledge of the background process.  For the purposes of discussion, we assume that we will model the background based on yields in neighboring kinematic bins; this method is usually called a ``sideband'' method because the background is under a ``foreground peak'' and the sidebands are kinematic bins to the left and the right of the peak.  We assume that the asymmetry in the sidebands is the same as that in the background under the peak; this is a strong assumption at the heart of any sideband method, and it should be verified if possible by studying the behavior of the background processes in data or in simulation.  We will call our estimate of the background yield, based on the sidebands, $Y_{SB}$.
\begin{equation}
Y^{\pm}_{SB}(\phi)=L^{\pm}\sigma_B[1+P^{\pm }A_B\sin(\pm\pi/2 -\phi)]
\label{eq:sideband_yield}
\end{equation}
The yield $Y_{SB}$ is a measured quantity, just like $Y_T$, since it is based on the yields in the sideband kinematic bins; some model is required to determine the estimate $Y_{SB}$ from the yields in those sideband bins, and the uncertainties in that model will be a source of systematic uncertainty in the final result.

We can now proceed to form the luminosity-normalized experimental asymmetry in the total yield,
\begin{eqnarray}
a_T(\phi) &=& \left.\left(\frac{Y^+_T(\phi)}{L^+}-\frac{Y^-_T(\phi)}{L^-}\right)\middle/\left(\frac{Y^+_T(\phi)}{L^+}+\frac{Y^-_T(\phi)}{L^-}\right)\right. \\
&=& \frac{(P^++P^-)(\sigma_F A_F\cos\phi +\sigma_B A_B \cos\phi)}{2[\sigma_F+\sigma_B]+(P^+-P^-)(\sigma_F A_F\cos\phi +\sigma_B A_B \cos\phi)} \\
&=& \frac{(P^++P^-)( A_F\cos\phi +f_B(\phi) A_B \cos\phi)}{2[1+f_B(\phi)]+(P^+-P^-)( A_F\cos\phi +f_B(\phi) A_B \cos\phi)}
\label{eq:total_asym}
\end{eqnarray}
where we have used the property that $\sin(\pm \pi/2-\phi)=\pm\cos\phi$.  The factor $f_B(\phi) = \sigma_B/\sigma_F$ is the background/foreground ratio, which may depend on the $\phi$ bin if the background process and/or the detector efficiency has some $\phi$-dependence independent of the spin-correlated one.  We determine the factor $f_B(\phi)$ from the ratio of the sideband-based esimate of the background yield $Y_{SB}(\phi)$ to the estimated foreground yield $Y_T(\phi)-Y_{SB}(\phi)$, summed over spin:
\begin{eqnarray}
y_R(\phi) &=& \left.\left(\frac{Y^+_{SB}(\phi)}{L^+}+\frac{Y^-_{SB}(\phi)}{L^-}\right)\middle/\left(\frac{Y^+_T(\phi)-Y_{SB}^+(\phi)}{L^+}+\frac{Y^-_T(\phi)-Y_{SB}^-(\phi)}{L^-}\right)\right. \\
&=& \frac{2\sigma_B+(P^+-P^-)\sigma_B A_B\cos\phi }{2\sigma_F+(P^+-P^-)\sigma_F A_F\cos\phi} \\
&=& f_B(\phi)\frac{2+(P^+-P^-) A_B\cos\phi }{2+(P^+-P^-) A_F\cos\phi}
\label{eq:yield_ratio}
\end{eqnarray}
Then we can find the foreground/background ratio,
\begin{eqnarray}
f_B(\phi) &=& y_R(\phi)\frac{1+\frac{1}{2}(P^+-P^-) A_F\cos\phi }{1+\frac{1}{2}(P^+-P^-) A_B\cos\phi} \\
&\approx& y_R(\phi)\left[ 1+\frac{1}{2}(P^+-P^-)(A_F-A_B)\cos\phi\right]
\label{eq:fb_ratio}
\end{eqnarray}
where we have made the assumption that $\left|\frac{1}{2}(P^+-P^-)A_B\cos\phi\right| \ll 1$.  If that is the case, then we may also safely conclude that $f_B(\phi)\approx y_R(\phi)$.  This assumption will only become a problem if the polarizations $P^+$ and $P^-$ are very different from one another, and we will check for bias produced by this assumption when we show tests of this method.

Having discovered the foreground/background ratio $f_B(\phi)$, we now turn our attention to the background asymmetry term $A_B\cos\phi$.  This is determined from the asymmetry in the sideband yields.
\begin{eqnarray}
a_{SB}(\phi) &=& \left.\left(\frac{Y^+_{SB}(\phi)}{L^+}-\frac{Y^-_{SB}(\phi)}{L^-}\right)\middle/\left(\frac{Y^+_{SB}(\phi)}{L^+}+\frac{Y^-_{SB}(\phi)}{L^-}\right)\right.\\
&=& \frac{(P^++P^-)\sigma_B A_B \cos\phi}{2\sigma_B+(P^+-P^-)\sigma_B A_B \cos\phi} \\
&=& \frac{(P^++P^-)A_B \cos\phi}{2+(P^+-P^-) A_B \cos\phi}
\label{eq:sideband_asym}
\end{eqnarray}
Then solving for $A_B\cos\phi$,
\begin{equation}
A_B\cos\phi = \frac{2a_{SB}(\phi)}{(P^++P^-)-(P^+-P^-)a_{SB}(\phi)}\equiv a'_{SB}(\phi)
\label{eq:background_asym}
\end{equation}
we can now substitute the histogram $a'_{SB}(\phi)$ wherever we see $A_B\cos\phi$.  The experimental asymmetry in the total yield may now be written as follows.
\begin{eqnarray}
a_T(\phi) &=&  \frac{(P^++P^-)[ A_F\cos\phi +y_R(\phi) a'_{SB}(\phi)]}{2[1+y_R(\phi)]+(P^+-P^-)[A_F\cos\phi +y_R(\phi) a'_{SB}(\phi)]}
\label{eq:corrected_total_asym}
\end{eqnarray}
From this we may solve for $A_F\cos\phi$, which is our goal.
\begin{equation}
A_F\cos\phi = \frac{a_T(\phi)\left[2[1+y_R(\phi)] +(P^+-P^-)y_R(\phi)a'_{SB}(\phi)\right]-(P^++P^-)y_R(\phi)a'_{SB}(\phi)}{(P^++P^-)-(P^+-P^-)a_T(\phi)}
\label{eq:final_histo}
\end{equation}
A simple fit of the right-hand-side of equation~\ref{eq:final_histo} to a cosine function yields the foreground asymmetry $A_F$.

\section{Unbinned Estimation of the Asymmetry:  without Unfolding}
\label{sec:unbinned_no_unfold}

A given scattering event $i$ will have particular values of polarization magnitude $P_i$, polarization direction $\phi_{{\rm pol},i}$, and azimuthal angle $\phi_i$.  The normalized probability distribution function ${\mathcal P}_i$ of such an event, based on the form of the azimuthal component of the angular distribution, would be 
\begin{equation}
    {\mathcal P}_i = \frac{1}{2\pi}\left[1+P_i A_N\sin(\phi_{{\rm pol},i}-\phi_i)\right].
    \label{eq:propability}
\end{equation}
Combining all of the events together we create the {\em likelihood} $\mathcal{L}$ that this collection of events could occur given a particular value of $A_N$.
\begin{equation}
{\mathcal L} = \prod_i \frac{1}{2\pi}[1+P_i A_N\sin(\phi_{{\rm pol},i}-\phi_i)]
\end{equation}
For numerical reasons it is more common to use the natural logarithm of the likelihood.
\begin{equation}
\ln{\mathcal L} = \sum_i \ln \left(\frac{1}{2\pi}[1+P_i A_N\sin(\phi_{{\rm pol},i}-\phi_i)]\right)
\end{equation}
The best value of the asymmetry, $\hat{A}_N$, based on our set of measurements, is determined by maximizing the log-likelihood.
\begin{equation}
\frac{d}{d A_N}\ln{\mathcal L} = \left.\sum_i \frac{d}{d A_N}\ln \left([1+P_i A_N\sin(\phi_{{\rm pol},i}-\phi_i)]\right)\right|_{\hat{A}_N}=0
\end{equation}
The uncertainty $\sigma_{\hat{A}_N}$ is found from the curvature of the likelihood at the extremum point.
\begin{equation}
    \sigma^2_{\hat{A}_N}=\left.\left(-\frac{d^2\ln {\mathcal L}}{dA^2_N}\right)^{-1}\right|_{\hat{A}_N} =\left(\sum_i P_i^2 \sin^2(\phi_{{\rm pol},i}-\phi_i)\right)^{-1}
    \label{eq:EAN}
\end{equation}

It may be necessary for each event to carry a different {\em weight}, $w^{\rm exp}_i$.  For example, if the set of events with spin up was associated with a different integrated luminosity than those with spin down, then the spin-up events should be weighted differently than the spin-down events, in proportion to the luminosity ratio between spin-up and spin-down datasets.  We use the superscript ``exp'' to denote these weights are determined by the experiment.  Then the likelihood would take this form.
\begin{equation}
    \ln {\mathcal L} = \sum_i w^{\rm exp}_i  \ln \left(\frac{1}{2\pi}[1+P_i A_N\sin(\phi_{{\rm pol},i}-\phi_i)] \right)
    \label{eq:logL}
\end{equation}
When weights are used then equation~\ref{eq:EAN} is not the correct one for the uncertainty.  Langenbruch~\cite{Langenbruch:2019nwe} has developed the correct form for the uncertainty when weights are used, and in our situation the result takes this form:
\begin{equation}
\sigma^2_{\hat{A}_N} = \frac{\sum_i \left(w^{\rm exp}_i \left. \frac{P\sin(\phi_{\rm pol}-\phi)}{1+P A_N \sin(\phi_{\rm pol}-\phi)}\right|_{\hat{A}_N}\right)^2}
{\left(\sum_i w^{\rm exp}_i \left. \frac{P^2\sin^2(\phi_{\rm pol}-\phi)}{[1+P A_N \sin(\phi_{\rm pol}-\phi)]^2}\right|_{\hat{A}_N}\right)^2}.
\label{eq:Lang}
\end{equation}
\smallskip
\subsection{Experimental Weights Correcting for Luminosity and/or Polarization Imbalances}
\label{sec:weights}

Let us recall what occurs in a binned analysis extracting a single-spin asymmetry. The number of events we might expect for a particular $\phi$ bin is:
\begin{equation}
Y^{\pm}(\phi)=L^{\pm}\sigma[1+P^{\pm }A_N\sin(\pm\pi/2 -\phi)]
\label{eq:yield}
\end{equation}
where $L$ is the integrated luminosity, $\sigma$ is the spin-averaged cross section, and the ``$\pm$'' superscript refers to the spin-up ($+$) or spin-down ($-$) set of events.  In a binned analysis, we normalize the yield histograms $Y^{\pm}$ according to the luminosity:  $Y_N^{\pm}(\phi)=Y^{\pm}(\phi)/L^{\pm}$ would be used for subsequent analysis.  In an unbinned analysis, we use event weights ($w^{\rm exp}$) to accomplish the same goal.  The form for these weights should satisfy certain constraints.  We want the weight to be inversely proportional to the luminosity; if one spin state has twice as much luminosity as the other, then we want to weight that spin-state half as much as the other.  The weight should be a number of order $\mathcal{O}(1)$ to avoid numerical problems.  If the two luminosities are the same, then we want the weights to be equal to unity.  A form satisfying those three constraints is:
\begin{equation}
    w^+ = \frac{L^+ + L^-}{2L^+}   ~~~~~~~~  w^- = \frac{L^+ + L^-}{2L^-}.
\end{equation}
What if the polarization magnitudes $P^{\pm}$ are not equal to each other? The weights must also take this into account.  In the expression for the yield, equation~\ref{eq:yield}, we see that the product $LPA_N$ occurs; to remove bias in the determination of $A_N$ we need to weight according to the product $LP$.  So the weights taking into account imbalances in both $L^{\pm}$ and $P^{\pm}$ are
\begin{equation}
    w^+ = \frac{L^+P^+ + L^-P^-}{2L^+P^+}   ~~~~~~~~  w^- = \frac{L^+P^+ + L^-P^-}{2L^-P^-}.
    \label{eq:weights}
\end{equation}
These will be the weights $w^{\rm exp}$ that we will use for spin-up and spin-down events.  We will demonstrate that these weights enable the log-likelihood analysis to return a correct estimate of $A_N$.

\subsection{Approximation of the Log-Likelihood for Small Asymmetries}

It is very common that the absolute value of the asymmetry $A_N$ is much smaller than unity.  In this circumstance we can make use of the expansion of $\ln(1+x)$ for small $x$.
\begin{equation}
    \ln(1+x)\approx x-\frac{1}{2}x^2+... ~~~~~~~ |x|<<1
\end{equation}
Then the log of the likelihood may be usefully approximated as follows.
\begin{equation}
    \ln {\mathcal L} = \sum_i w^{\rm exp}_i \frac{1}{2\pi}\left[P_i A_N\sin(\phi_{{\rm pol},i}-\phi_i) -\frac{1}{2}P^2_i A^2_N\sin^2(\phi_{{\rm pol},i}-\phi_i)\right]
\end{equation}
The optimum value of $A_N$ can then by found deterministically by finding the extremum of the likelihood.
\begin{equation}
    \frac{d\ln {\mathcal L}}{dA_N} =\frac{1}{2\pi} \sum_i w^{\rm exp}_i \left[P_i \sin(\phi_{{\rm pol},i}-\phi_i) -A_N P^2_i \sin^2(\phi_{{\rm pol},i}-\phi_i)\right]=0
\end{equation}
\begin{equation}
\Rightarrow \hat{A}_N = \frac{\sum_i w^{\rm exp}_i P_i \sin(\phi_{{\rm pol},i}-\phi_i)}{\sum_i w^{\rm exp}_i P_i^2 \sin^2(\phi_{{\rm pol},i}-\phi_i)}
\label{eq:AN}
\end{equation}
The uncertainty $\sigma_{\hat{A}_N}$ is found using equation~\ref{eq:Lang}.

\subsection{Unbinned Log-Likelihood when a Background must be Subtracted}

It is very common that a background must be removed in order to determine the true yield for the events of interest.  The background might be due to another physics process, or combinatorics, or some other issue.  We will not make any assumptions about the cause of the background here; we will simply assume that some model has been used to estimate the background events.

As mentioned earlier, for the purposes of discussion, we will consider the case when the background is under a ``peak'' in a spectrum, and we use ``sidebands'' to the left and the right of the peak to estimate the background events under the peak.  A model of some sort must be used to determine the estimate of the background events by using the sideband events; we do not make any assumptions about that model here, except that we assume that it has been done correctly.  Of course, the parameters of that model will need to be varied to determine the systematic uncertainty associated with that model, and the machinery developed here can be used for that determination.

Since the background is to be subtracted then in the weighted log-likelihood method we give the background events a negative weight.  We will demonstrate the success of this very simple prescription.

\section{Generating Events to Test the Method}
\label{sec:gen_events}

To test the binned analysis and the unbinned log-likelihood technique, we generate events based on our model. 
Assume a typical situation in an experiment with a spin-polarized target where events are measured for two spin alignments: spin ``up'' and spin ``down'' are represented respectively by $\phi_{{\rm pol}} = +\frac{\pi}{2}$ and $-\frac{\pi}{2}$. A recorded event $i$ will contain a spin orientation, an azimuthal angle, and a measured polarization value: $(\phi_{{\rm pol},i} , \phi_i , P_i)$.
The integrated luminosities for the spin-up ($L^+$) and spin-down ($L^-$) data sets are known.

\begin{itemize}
    \item $\phi_{i}$ - drawn randomly from a uniform distribution within the range $[-\pi,\pi]$.
    \item $\phi_{{\rm pol},i}$ - assigned randomly to either $-\frac{\pi}{2}$ or $+\frac{\pi}{2}$, according to the relative luminosities associated with the two spin states.
    \item $P_i$ - these might change with time; for the moment we will use fixed values for the spin-up ($P^+$) and spin-down ($P^-$) polarizations, which should not be assumed to be equal.
\end{itemize}

The experiment will have an acceptance and/or efficiency $e(\phi)$ for detecting events as a function of $\phi$.  This needs to be taken into account when generating simulated events.  However, in our analysis of those events we will not assume we know what the form of the efficiency is.
In our tests we used three simple forms for the efficiency.
\begin{itemize}
    \item $e(\phi)=1$ ~~~~~ a perfect detector
    \item $e(\phi)=\frac{1}{2}\left[1+\frac{1}{2}\sin\phi\right]$ ~~~~~ containing a $\sin\phi$ dependence
    \item $e(\phi)=\frac{1}{2}\left[1+\frac{1}{2}\cos\phi\right]$ ~~~~~ containing a $\cos\phi$ dependence
\end{itemize}
The third form for the efficiency, containing a $\cos\phi$ term, is critical because this contains the same azimuthal dependence as our physics signal:  $\sin(\pm \pi/2-\phi)=\pm\cos\phi$.  Therefore this efficiency form is the most stringent test of our analysis scheme.

Now the azimuthal modulation needs to be embedded into this set of events. This was done by calculating a weight $W_i$ expressed by
\begin{equation}
    W_i = \left[1+ P_i A_N \sin(\phi_{{\rm pol},i}-\phi_i)\right]e(\phi)
    \label{eqWi}
\end{equation}
and comparing it with a random number generated uniformly in the range $[0,2]$; if the random number was less than the weight $W_i$ then we kept the event.  In this way the asymmetry $A_N$, the polarization $P$, and the efficiency $e(\phi)$ all play a role in determining if an event is kept or not.

If a background is to be included, we need to revise our method of generating test events.  There needs to be a ``truth'' yield and asymmetry for the events of interest (the foreground), a truth yield and asymmetry for the background, and a truth yield and asymmetry for the sideband-based model of the background.  We assume the same asymmetry for the sideband events as for the true background.  We need also to assume an integral foreground/background ratio, which is used to determine (randomly) which kind of event is to be generated during each generation step.  The lists of truth foreground and background events must be combined into a single ``total'' list of events, since we cannot distinguish them from each other in the actual experiment; they all contribute to the total yield in the region of interest of the peak.  The sideband events remain in their own event list. 

A set of four standard datasets were developed, for testing all the analysis methods.  These are defined in table~\ref{tab:setups}.  Other datasets were also created for various purposes and these are described separately as needed.

\begin{table}[ht]
\begin{center}
\caption{List of parameters for each data set used for testing the unfolding.  The tests, from left to right, are increasing more complex.  The label ``Pol/lumi Imbalance'' means that the polarization magnitudes and integrated luminosities are not equal between the two spin states of the target.}
\label{tab:setups}
\begin{tabular}{||L|L|L|L|L||}  \hline 
 Test Name & ``Simple'' & ``Background Asymmetry'' &
 ``Pol/lumi Imbalance" & ``Cosine Efficiency'' \\ \hline
 \hline 
Number of events & 200,000 & 200,000 & 200,000 & 200,000 \\ \hline
Background to Foreground ratio & 0.2 & 0.2 & 0.2 & 0.2 \\ \hline 
Foreground Asymmetry & 0.2 & 0.2 & 0.2 & 0.2 \\ \hline 
Background Asymmetry & 0.0 & -0.1 & -0.1 & -0.1 \\ \hline 
Polarizations& $P^+=1.0$ $P^-=1.0$ & $P^+=1.0$ $P^-=1.0$ & $P^+=0.9$ $P^-=0.7$ & $P^+=0.9$ $P^-=0.7$ \\ \hline 
Luminosity Ratio up:down & 1:1 & 1:1 & 3:7 & 3:7 \\ \hline 
Efficiency & 1 & 1 & 1 & $\frac{1}{2}[1+\frac{1}{2}\cos\phi]$ \\ \hline 
\end{tabular}
\end{center}
\end{table}

\section{Results of Test of the Method}

\subsection{Binned Analysis}

For the binned analysis, using the lists of events described in Section~\ref{sec:gen_events}, we create histograms for the total yield $Y_T^\pm(\phi)$ and for the sideband-based estimate of the background $Y_{SB}^\pm(\phi)$, and then proceed with the analysis described in Section~\ref{sec:binned_no_unfold}.

\begin{table}[ht]
\centering
\caption{Listing of results when each individual test was repeated 1000 times.  The ``Mean $A_N$'' is the mean of the histogram of 1000 test results for the foreground asymmetry, $\sigma_{1000}$ is the width of that histogram, while $\sigma_1$ is the statistical uncertainty associated with a single result.  The Analysis Methods are: (a) Binned (Section~\ref{sec:binned_no_unfold}), and (b) Unbinned Log-Likelihood (Section~\ref{sec:unbinned_no_unfold}).}
\label{tab:thousand}
\begin{tabular}{||L|L|L|L|L||}  \hline 
 & Analysis Method & Mean $A_N$ & $\sigma_{1000}$ & $\sigma_1$ \\
\hline \hline
\multirow{2}{4em}{``Simple''}& (a) &0.1977&0.0044&0.0044\\
 & (b) &0.1999&0.0044&0.0044\\
 \hline
\multirow{2}{4em}{``Background Asymmetry''}& (a) &0.1976&0.0046&0.0044\\
 & (b) &0.1999&0.0046&0.0044\\
 \hline
\multirow{2}{4em}{``Pol/lumi Imbalance''}& (a) &0.1975&0.0061&0.0060\\
 & (b) &0.1999&0.0061&0.0060\\
 \hline
\multirow{2}{4em}{``Cosine Efficiency''}& (a) &0.1971&0.0057&0.0060\\
 & (b) &0.1968&0.0056&0.0060\\
 \hline
\end{tabular}
\end{table}

\subsection{Unbinned Log-likelihood Analysis}

In the weighted log-likelihood analysis, the events in the total list (true foreground plus true background) get a positive weight of the form equation~\ref{eq:weights}, and the events in the sideband list get a similar negative weight.  Then we can either maximize the $\ln {\mathcal L}$ or use equation~\ref{eq:AN} to find the optimal foreground $A_N$.  We continue to use equation~\ref{eq:Lang} for the uncertainty in $A_N$ in either case.  

We first use the standard datasets defined in table~\ref{tab:setups} to illustrate the statistical properties of both the binned (Section~\ref{sec:binned_no_unfold}) and unbinned (Section~\ref{sec:unbinned_no_unfold}) methods of analysis.  Each method was used 1000 times, the results were histogramed, and the mean and width of the histogram were measured.  The results are shown in table~\ref{tab:thousand}.
We see that the results are distributed with a width consistent with the uncertainty of a single measurement, and the mean $A_N$ value is within one standard deviation of the injected value of 0.2. This demonstrates the correctness of these two methods. Note that the statistical significance of the result suffers when the polarizations and integrated luminosities for the two spin states are imbalanced, as expected; most experiments try to keep the polarization magnitudes the same and the integrated luminosities the same, for this reason.  Even so, it is well to note that when the polarizations and luminosities are unbalanced, there is no evidence of any bias in the result; this justifies our use of the approximation $f_B(\phi)\approx y_R(\phi)$ in section~\ref{sec:binned_no_unfold}.

With our test apparatus in hand, we can demonstrate many properties of the log-likelihood technique in regards to the extraction of the asymmetry $A_N$. For the moment, we will not include any background events in our test.
\begin{itemize}
\item If the efficiency does not contain a term that looks like the physics signal (in our case, the $\cos\phi$ term), then it is not necessary to flip spin.  The log-likelihood technique will return the correct value of $A_N$.  This is illustrated by a few examples in table~\ref{tab:simple}. 
\begin{table}[h!]
\begin{center}
 \caption{Extracted asymmetry using Equation~\ref{eq:AN} under three different detector efficiency models $e(\phi)$, and two different choices for the relative luminosities, 100\% spin up or $50-50$\% spin up-down.  We have used a polarization of $P=1$ for all events.  The embedded asymmetry was $A_N=0.2$. There is no background in this set of tests.  Each simulated data sample contained 500,000 events. Note how the technique fails for the 100\% spin-up case when the efficiency contains a $\cos\phi$ term.}
 \label{tab:simple}
 \begin{tabular}{||c| c| c||} 
 \hline Efficiency Model 
   & 100\% spin up & $50-50$\% spin up-down \\ [0.5ex] 
 \hline\hline
 Perfect detector : $e(\phi_i)=1.0$ & $0.199\pm 0.031$ & $0.201\pm 0.002$ \\ [0.5ex]
 \hline
 $e(\phi)=\frac{1}{2}\left[1+\frac{1}{2}\sin\phi\right]$ & $0.199\pm 0.031$ & $0.200\pm 0.002$ \\ [0.5ex]
 \hline
$e(\phi)=\frac{1}{2}\left[1+\frac{1}{2}\cos\phi\right]$ & $0.650 \pm 0.002$ & $0.199\pm 0.002$ \\ [0.5ex]
 \hline
\end{tabular}
\end{center}
\end{table}
The technique fails when the efficiency contains a $\cos\phi$ term and there is no spin-flipping.  It is worth noting that the statistical significance is greatly improved by flipping the spin.  And, of course, in practice one can never be absolutely sure the acceptance and/or efficiency does not contain a component that imitates the physics signal. For these reasons, every experiment involving a polarized beam or target will reverse the spin direction from time to time.
\item If the luminosities for the two spin-states are equal, and also the polarization magnitudes $P^+$ and $P^-$ are equal, then the weights in equation~\ref{eq:weights} are both equal to 1, and then the log-likelihood method will work for any efficiency model. This is illustrated with a few examples in table~\ref{tab:balanced}.
\begin{table}[h!]
\begin{center}
 \caption{Extracted asymmetry using equation~\ref{eq:AN} under three different detector efficiency models $e(\phi)$, with equal luminosities for each spin state, and equal polarization magnitudes $P^\pm =0.9$.  The embedded asymmetry was $A_N=0.1$.  Each simulated data sample contained 50,000 events. }
 \label{tab:balanced}
 \begin{tabular}{||c| c||} 
 \hline Efficiency Model 
   & Extracted asymmetry\\
 \hline\hline
 Perfect detector : $e(\phi_i)=1.0$ & $0.103 \pm 0.007$ \\
 \hline
 $e(\phi)=\frac{1}{2}\left[1+\frac{1}{2}\sin\phi\right]$ & $0.106\pm 0.007$  \\
 \hline
$e(\phi)=\frac{1}{2}\left[1+\frac{1}{2}\cos\phi\right]$ & $0.104 \pm 0.007$\\
 \hline
\end{tabular}
\end{center}
\end{table}
\item If the luminosities and/or polarization magnitudes for the two spin-states are unequal ($L^+\neq L^-$ and/or $P^+ \neq P^-$), then the weights in equation~\ref{eq:weights} must be used to be sure that the log-likelihood method will work for any efficiency model. This is illustrated with many examples in table~\ref{tab:unbalanced}.  Note that the statistical significance suffers if the imbalance in luminosity or polarization is very large.
\begin{table}[h!]
\begin{center}
 \caption{Extracted asymmetry using equation~\ref{eq:AN} under three different detector efficiency models $e(\phi)$, with unequal luminosities for each spin state, and unequal polarization magnitudes. The embedded asymmetry was $A_N=0.1$.  Each simulated data sample contained 50,000 events. }
 \label{tab:unbalanced}
 \begin{tabular}{||c| c| c| c|c||} 
 \hline Efficiency Model & Relative Luminosity & $P^+$ & $P^-$ 
   & Extracted asymmetry\\
 \hline\hline
 $e(\phi_i)=1.0$ & 40:60 up:down & 0.9 & 0.7 & $0.105 \pm 0.008$ \\
 \hline
 $e(\phi_i)=1.0$ & 30:70 up:down & 0.6 & 0.7 & $0.109 \pm 0.011$ \\
 \hline
 $e(\phi_i)=1.0$ & 20:80 up:down & 0.9 & 0.4 & $0.110 \pm 0.012$ \\
 \hline
 $e(\phi)=\frac{1}{2}\left[1+\frac{1}{2}\sin\phi\right]$ & 40:60 up:down & 0.9 & 0.7 & $0.105\pm 0.008$  \\
 \hline
 $e(\phi)=\frac{1}{2}\left[1+\frac{1}{2}\sin\phi\right]$ & 30:70 up:down & 0.6 & 0.7 & $0.103\pm 0.011$  \\
 \hline
 $e(\phi)=\frac{1}{2}\left[1+\frac{1}{2}\sin\phi\right]$ & 20:80 up:down & 0.9 & 0.4 & $0.104\pm 0.012$  \\
 \hline
$e(\phi)=\frac{1}{2}\left[1+\frac{1}{2}\cos\phi\right]$ & 40:60 up:down & 0.9 & 0.7 & $0.105 \pm 0.008$\\
 \hline
 $e(\phi)=\frac{1}{2}\left[1+\frac{1}{2}\cos\phi\right]$ & 30:70 up:down & 0.6 & 0.7 & $0.103 \pm 0.010$\\
 \hline
 $e(\phi)=\frac{1}{2}\left[1+\frac{1}{2}\cos\phi\right]$ & 20:80 up:down & 0.9 & 0.4 & $0.099 \pm 0.012$\\
 \hline
\end{tabular}
\end{center}
\end{table}
\end{itemize}

In some experiments, the magnitude of polarization may vary as a function of time; for example, radiation damage to the target material may cause a reduction in polarization.
To consider this possibility, we introduced a uniformly varying polarization within a range, for example $[0.75, 0.95]$, and the range might be different for spin-up and spin-down events. The weights in equation~\ref{eq:weights} make use of the average polarization values.  The extracted asymmetries (see table~\ref{tab:ranges}) are always within one standard deviation of the injected value. A varying polarization does not diminish the extraction power of the technique.  Once again, when the imbalance in luminosity or polarization is large, then the statistical significance suffers.

\begin{table}[ht]
\begin{center}
    \caption{Extracted asymmetries with from a data set using uniformly varying polarizations. The efficiency model in all cases was $e(\phi)=\frac{1}{2}\left[1+\frac{1}{2}\cos\phi\right]$. The embedded asymmetry was $A_N=0.2$.  Each simulated data set contained 50,000 events.}
 \label{tab:ranges}
 \begin{tabular}{||c| c| c||} 
 \hline Polarization Ranges & Relative Luminosity & Extracted Asymmetry \\
 \hline\hline
 [0.75,0.95] up ~~~ [0.70,0.90] down & 50:50 up:down & $0.203 \pm 0.008$ \\
 \hline
 [0.75,0.95] up ~~~ [0.60,0.80] down & 70:30 up:down & $0.205 \pm 0.009$ \\
 \hline
 [0.75,0.95] up ~~~ [0.60,0.80] down & 30:70 up:down & $0.196 \pm 0.009$ \\
 \hline
[0.75,0.95] up ~~~ [0.75,0.95] down & 50:50 up:down & $0.202 \pm 0.007$ \\
 \hline\hline
\end{tabular}
\end{center}
\end{table}

To make an estimate of the systematic error introduced by terminating the power series of $\ln(1+x)$ at the second order, we compared the results calculated using our deterministic formula (equation~\ref{eq:AN}) with the results obtained by maximizing the likelihood expressed in equation~\ref{eq:logL} as a function of increasing $A_N$. With a dataset of size $10^5$, the bias arising from the termination of the power series is significant only in the fourth decimal place, see table \ref{tab:power}. This is dominated by the statistical error at this size dataset. 

\begin{table}[ht]
\begin{center}
    \caption{Extracted asymmetries using both the maximization of $\ln {\mathcal L}$ and using  equation~\ref{eq:AN}. The efficiency model in all cases was $e(\phi)=\frac{1}{2}\left[1+\frac{1}{2}\cos\phi\right]$. Tests were carried out with a uniformly varying polarization within $[0.75, 0.95]$ and equal luminosities for spin-up and spin-down events, for a dataset of size $10^5$.}
 \label{tab:power}
 \begin{tabular}{||c|c|c|c||} 
 \hline 
Embedded Asymmetry $A_N$ & $\ln {\mathcal L}$ maximization & Using eq.~\ref{eq:AN} & Uncertainty \\
 \hline\hline
0.1 & 0.0950 & 0.0950 & 0.0052 \\
 \hline
0.2 & 0.1959 & 0.1960 & 0.0052 \\
 \hline
0.3 & 0.2966 & 0.2967 & 0.0051 \\
 \hline
0.4 & 0.3981 & 0.3981 & 0.0050 \\
 \hline
0.5 & 0.5020 & 0.5022 & 0.0049 \\
 \hline
0.6 & 0.6016 & 0.6018 & 0.0047 \\
 \hline
0.7 & 0.7033 & 0.7038 & 0.0044 \\
 \hline
0.8 & 0.8025 & 0.8027 & 0.0041 \\
 \hline
\end{tabular}
\end{center}
\end{table}

Now we introduce a background with an asymmetry $A_B$ into the the events.  table~\ref{tab:background} displays the results of a set of increasingly complex tests; in every case the returned value of $A_N$ is within one standard deviation of the embedded value.

\begin{table}[ht]
\begin{center}
\caption{Extracted values of the asymmetry when background has been subtracted, under a variety of conditions. The ``number of events'' is summed over foreground, background, and sideband events.}
\label{tab:background}
\begin{tabular}{||L|L|L|L|L||}  \hline 
 & Simple Test & Add background asymmetry &
 Add polarization and luminosity imbalance & Add an efficiency \\ \hline 
Number of events & 50,000 & 50,000 & 50,000 & 50,000 \\ \hline
Background to Foreground ratio & 0.2 & 0.2 & 0.2 & 0.2 \\ \hline 
Foreground Asymmetry & 0.2 & 0.2 & 0.2 & 0.2 \\ \hline 
Background Asymmetry & 0.0 & 0.1 & 0.1 & 0.1 \\ \hline 
Polarization Range: up & [0.75,0.95] & [0.75,0.95] & [0.75,0.95] & [0.75,0.95] \\ \hline 
Polarization Range: down & [0.75,0.95] & [0.75,0.95] & [0.65,0.85] & [0.65,0.85] \\ \hline 
Luminosity Ratio up:down & 1:1 & 1:1 & 3:7 & 3:7 \\ \hline 
Efficiency & 1 & 1 & 1 & $\frac{1}{2}[1+\frac{1}{2}\cos\phi]$ \\ \hline 
\hline
Extracted Foreground $A_N$ & $0.2046 \pm 0.0103$ & $0.2017 \pm 0.0103$ & $0.2059 \pm 0.0119$ & $0.1911 \pm 0.0118$ \\
\hline
\end{tabular}
\end{center}
\end{table}

\section{Unfolding: Binned and Unbinned}
\label{sec:unfold}

The technique described in the previous sections is appropriate when the reconstruction of the azimuthal angle $\phi$ (along with other kinematic observables) is of sufficient quality to not have any appreciable affect on the determination of the asymmetry.  Unfortunately this is often not the case; it is often necessary to ``unfold'' the distortions in the reconstructed kinematic variables in order to extract good estimates of the physics quantities of interest.  In a binned analysis technique, there are many unfolding techniques available, for example Wiener-SVD~\cite{Tang:2017rob} and iterative Bayesian~\cite{DAGOSTINI1995487}. We will explore also unfolding techniques in the context of an unbinned maximum likelihood method.

\subsection{Likelihood Ratio Estimation Using Event Classification}

By way of introduction, it is useful to introduce some elementary concepts regarding unfolding and the associated reweighting of events.  The likelihood ratio can be used to map one probability distribution to another. In a binned scenario, the likelihood ratio can be estimated simply as the ratio between the target and source histograms. However, in the unbinned case, analytically formulating the likelihood ratio is often a challenging task. A common approach is to use an \textit{ad hoc} function to approximate the likelihood ratio, but this method can introduce large systematic uncertainties. To overcome this challenge, binary classifiers can be employed to approximate the likelihood ratios.

To demonstrate this method, we utilize the probability distribution given in equation~\ref{eq:propability}. We define $p(\phi \vert A_{N} = 0.0)$ as our source distribution and $p(\phi \vert A_{N} = 0.2)$ as the target distribution. The objective is to estimate the likelihood ratio $r(\phi) = \dfrac{p(\phi \vert A_{N} = 0.2)}{p(\phi \vert A_{N} = 0.0)}$. Let $s(\phi)$ be a binary classifier trained to distinguish samples drawn from $\phi \sim p(\phi \vert A_{N} = 0.0)$ (labeled $y = 0$) and $\phi \sim p(\phi \vert A_{N} = 0.2)$ (labeled $y = 1$). The likelihood ratio estimator can then be defined as;
\begin{equation}
\hat{r}(\phi) = \dfrac{s(\phi)}{1 - s(\phi)}
\label{likelihood}
\end{equation}
This technique is commonly referred to as the ``likelihood-ratio trick''~\cite{Andreassen:2019cjw, Cranmer:2015bka, Andreassen:2019nnm, Rizvi:2023mws}. This method calculates event weights that map $p(\phi \vert A_{N} = 0.0) \rightarrow p(\phi \vert A_{N} = 0.2)$.

Events were generated as described in Section~\ref{sec:gen_events}, with a luminosity ratio of 60:40 and a polarization ratio of 0.9:0.7 for the spin-up and spin-down configurations, respectively. For simplicity, we assume $\epsilon(\phi) = 1$. We employ a feed-forward neural network (FNN) consisting of three hidden layers, each containing 64 neurons. The hidden layers utilize the rectified linear unit (ReLU) activation function, while the output layer uses the Sigmoid activation function, $\sigma(x) = \dfrac{1}{1 + e^{-x}}$.

\begin{figure}[htbp]
    \centering
    \begin{subfigure}[b]{0.45\textwidth}
        \centering
        \includegraphics[width=\textwidth]{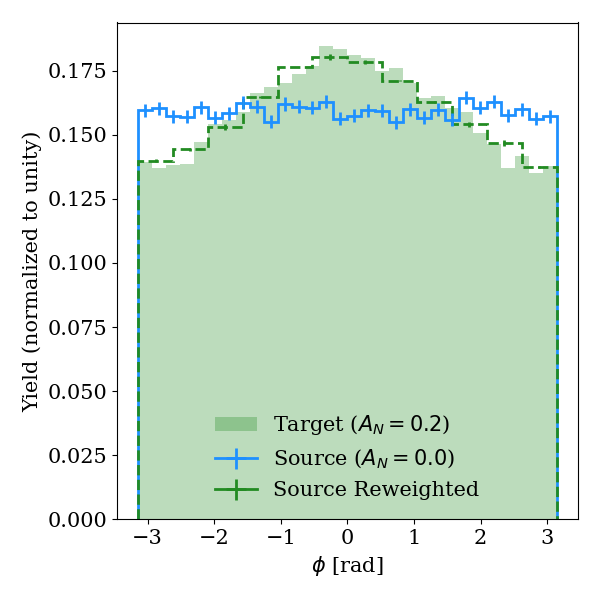}
        \caption{Spin up configuration.}
    \end{subfigure}
    \hfill
    \begin{subfigure}[b]{0.45\textwidth}
        \centering
        \includegraphics[width=\textwidth]{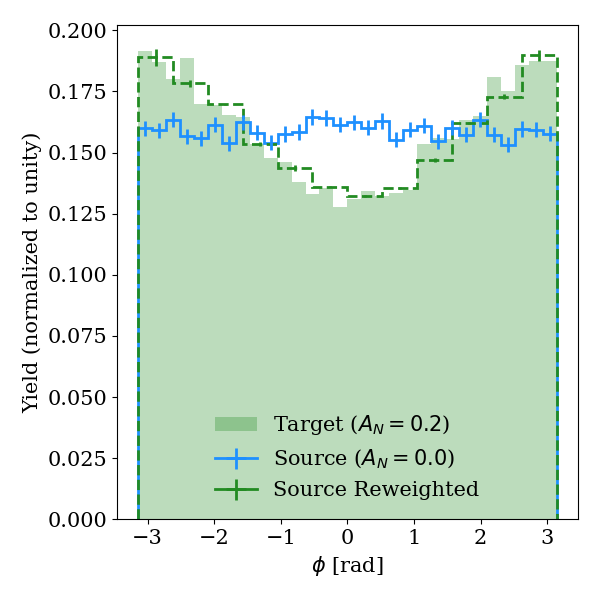}
        \caption{Spin down configuration.}
    \end{subfigure}
    \caption{Reweighted distributions using the likelihood ratio method. The blue line represents the source distribution, the filled area indicates the target distribution, and the dashed green line shows the reweighted source distribution with weights calculated using equation~\ref{likelihood}.}\label{reweighted}
\end{figure}

The \verb|Adam| optimizer was used for training with an initial learning rate of $1 \times 10^{-4}$, utilizing the \verb|ReduceLROnPlateau| scheduler. The batch size was set to 2048. The binary classifier was trained for 30 epochs, saving the model state with the lowest validation loss. To estimate uncertainties, the training was performed for 100 runs, resampling the source events with replacement in each run. The mean value of the bin contents across these 100 runs is taken as the final bin content, and the standard deviation across the runs is assigned as the bin error. The reweighted distributions are presented in Figure~\ref{reweighted}.

\begin{figure}[htbp]
    \centering
    \begin{subfigure}[b]{0.45\textwidth}
        \centering
        \includegraphics[width=\textwidth]{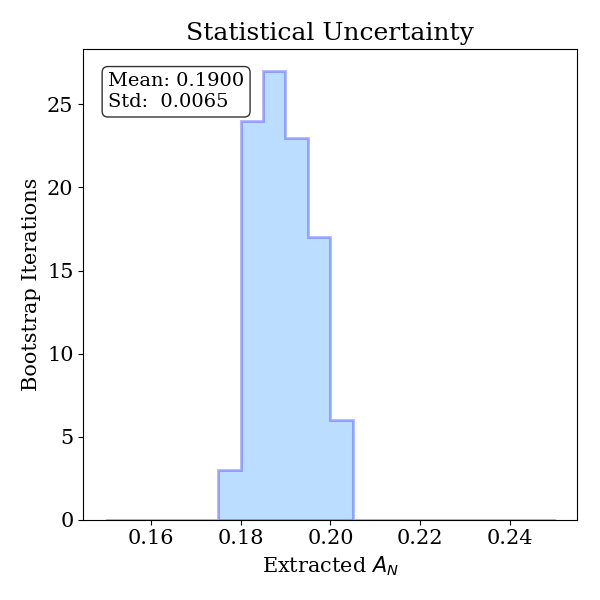}
        \caption{Extracted $A_{N}$ values across 100 runs with resampled source events.}\label{statisticalUncertainty}
    \end{subfigure}
    \hfill
    \begin{subfigure}[b]{0.45\textwidth}
        \centering
        \includegraphics[width=\textwidth]{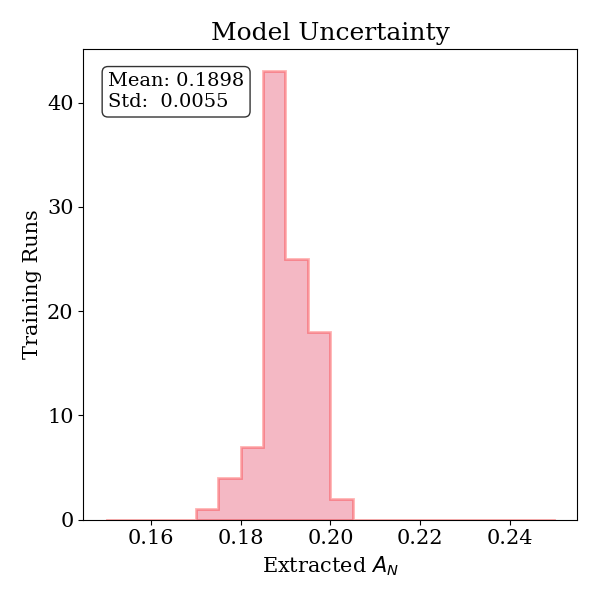}
        \caption{Extracted $A_{N}$ values across 100 runs with different random initializations.}\label{modelUncertainty}
    \end{subfigure}
    \caption{Statistical and model uncertainties.}
\end{figure}

We use equation~\ref{eq:AN} to extract the $A_{N}$ value from the reweighted source distribution. Since the source events are resampled in each run, the standard deviation across these 100 runs represents the statistical uncertainty of the extraction (Figure~\ref{statisticalUncertainty}). To account for the random initialization of the FNN and stochasticity in batch selection, we train the neural network for 100 separate runs with different random initializations and interpret the standard deviation across these runs as the model uncertainty (Figure~\ref{modelUncertainty}).~\footnote{GitHub: \url{https://github.com/dinupa1/NeuralLikelihoodRatio}}

This likelihood ratio estimation method can be extended to the unfolding problem in particle and nuclear physics. The OmniFold algorithm~\cite{Andreassen:2019cjw, Huang:2025ziq} uses this likelihood ratio method to iteratively reweight the source and target distributions at both the reconstruction level and the truth level.

\subsection{Unbinned Unfolding with OmniFold}

In this manuscript, three methods of asymmetry extraction involving unfolding will be illustrated.
\begin{enumerate}
\item[(1)] Unfolding of binned data, followed by extraction of the asymmetry using the technique described in Section~\ref{sec:binned_no_unfold}
\item[(2)] Unfolding of the unbinned data, followed by binning of the data and extraction of the asymmetry as in Section~\ref{sec:binned_no_unfold}
\item[(3)] Unfolding of the unbinned data, followed by extraction of the asymmetry using an unbinned likelihood optimization as in Section~\ref{sec:unbinned_no_unfold}
\end{enumerate}
For the unfolding of both binned and unbinned data, we will use a software package developed by Milton {\em et al.}~\cite{Milton:2025mug} available at \url{https://github.com/rymilton/unbinned_unfolding}.   The unbinned unfolding technique they have used is OmniFold~\cite{Andreassen:2019cjw}, a relatively recent development which offers simultaneous unbinned unfolding of multiple observables.  They have implemented the OmniFold technique using boosted decision trees (using the scikit-learn~\cite{scikit-learn} machine-leaning library) and packaged it into the commonly-used RooUnfold~\cite{Brenner:2019lmf} framework.  The ability to unfold binned data is also included in the package; the binned unfolding technique they use is based on the iterative Bayesian unfolding already found in RooUnfold. 

\begin{figure}[t]
    \centering
    \includegraphics[scale=0.7]{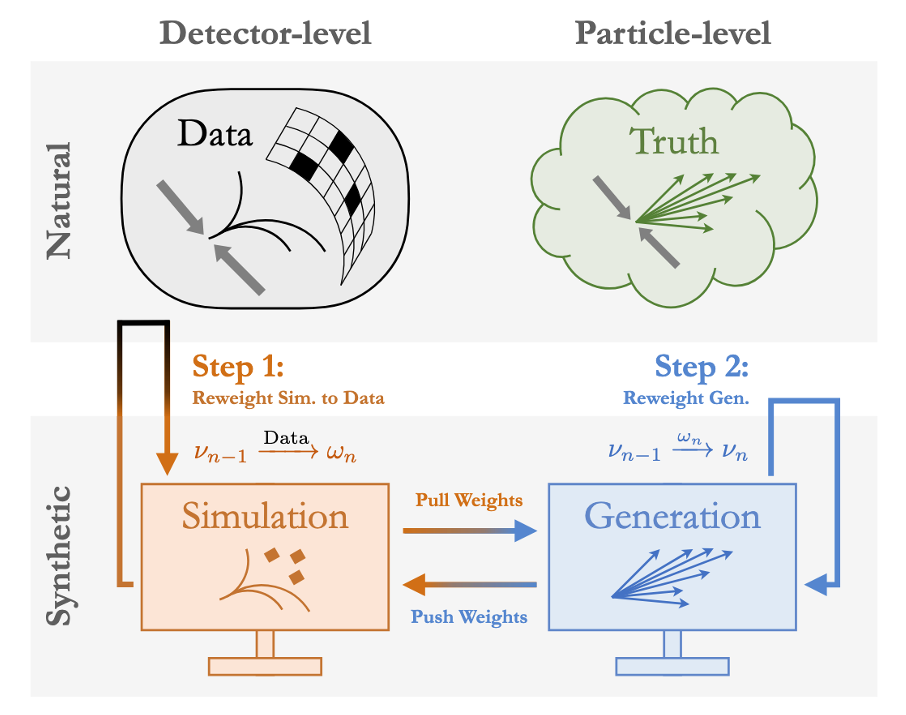}
    \caption{Diagram illustrating the OmniFold technique.  Taken from \cite{Andreassen:2019cjw}.}
    \label{fig:OmniFold}
\end{figure}

To explain the testing of this package, it is useful to briefly review the OmniFold technique and describe our work using the same language.  A diagram, taken from \cite{Andreassen:2019cjw} and reproduced in Figure~\ref{fig:OmniFold}, is useful for the explanation of OmniFold.  There are four datasets of interest.
\begin{itemize}
    \item  Particle-level {\bf Truth}: This is the information we seek to estimate from the data.  
    \item Detector-level {\bf Data}: This is what we measure with the detector.
    \item Particle-level {\bf Generation}:  Using a model and an event generator, this is our estimate for {\bf Truth}.  We will adjust this estimate using weights.
    \item Detector-level {\bf Simulation}:  The {\bf Generation} is passed through a simulation of the detector response to produce the {\bf Simulation}.
\end{itemize}
The Data and Simulation are compared and the Simulation events are reweighted to better match the Data.  These weights are ``pulled'' to the Generation side and used to induce a set of weights on the Generation events. Then the initial and revised Generation events are compared to create a new set of weights, which are ``pushed'' to the Simulation side, producing a new Simulation.  This process is iterated.  At the end, it is the reweighted Generation events that are used for final analysis, not the original Data.  This is very important to have in mind, since it means the statistical significance of the Generation must be similar to that of the Data, otherwise the statistical uncertainty in the final results will be wrongly calculated.

The artificial data used to test the unfolding are generated in the same way as described in Section~\ref{sec:gen_events}, except that we smear the $\phi$ angle of the detected particle to imitate the effects of track reconstruction.  We used a simple Gaussian smearing; we draw a random number from a Gaussian distribution of mean 0.0 and width $\sigma_{\rm smear}$, and added this number to the truth value of $\phi$.  We used two different values of $\sigma_{\rm smear}$ in our tests:  $\sigma_{\rm smear}=0.45$ radians which we will refer to as the ``weak smearing'' case, and $\sigma_{\rm smear}=0.90$ radians which we will refer to ``strong smearing''.  To be more specific, we describe the origin of each of the four datasets in the OmniFold language.
\begin{itemize}
    \item {\bf Truth}: We generate the Truth data using particular values of the asymmetries $A_F$ and $A_B$, some value for the background/foreground ratio, some distribution of polarizations, and a ratio of spin-up and spin-down luminosities.
    \item {\bf Data}: We smear the $\phi$ angle in Truth to produce the Data.
    \item {\bf Generation}:  This is independently produced in the same way as Truth, but with the asymmetries $A_F$ and $A_B$ set to 0.0.
    \item {\bf Simulation}: We smear the $\phi$ angle in Generation to produce the Simulation.
\end{itemize}

To prepare a test sample, it is necessary to specify a number of test parameters; the number of events, the magnitude of polarization, etc.  Table~\ref{tab:setups} lists increasingly complicated sets of parameters used in our testing.

\begin{figure}[ht]
\centering
Convergence test with weak smearing\\
\includegraphics[scale=0.4]{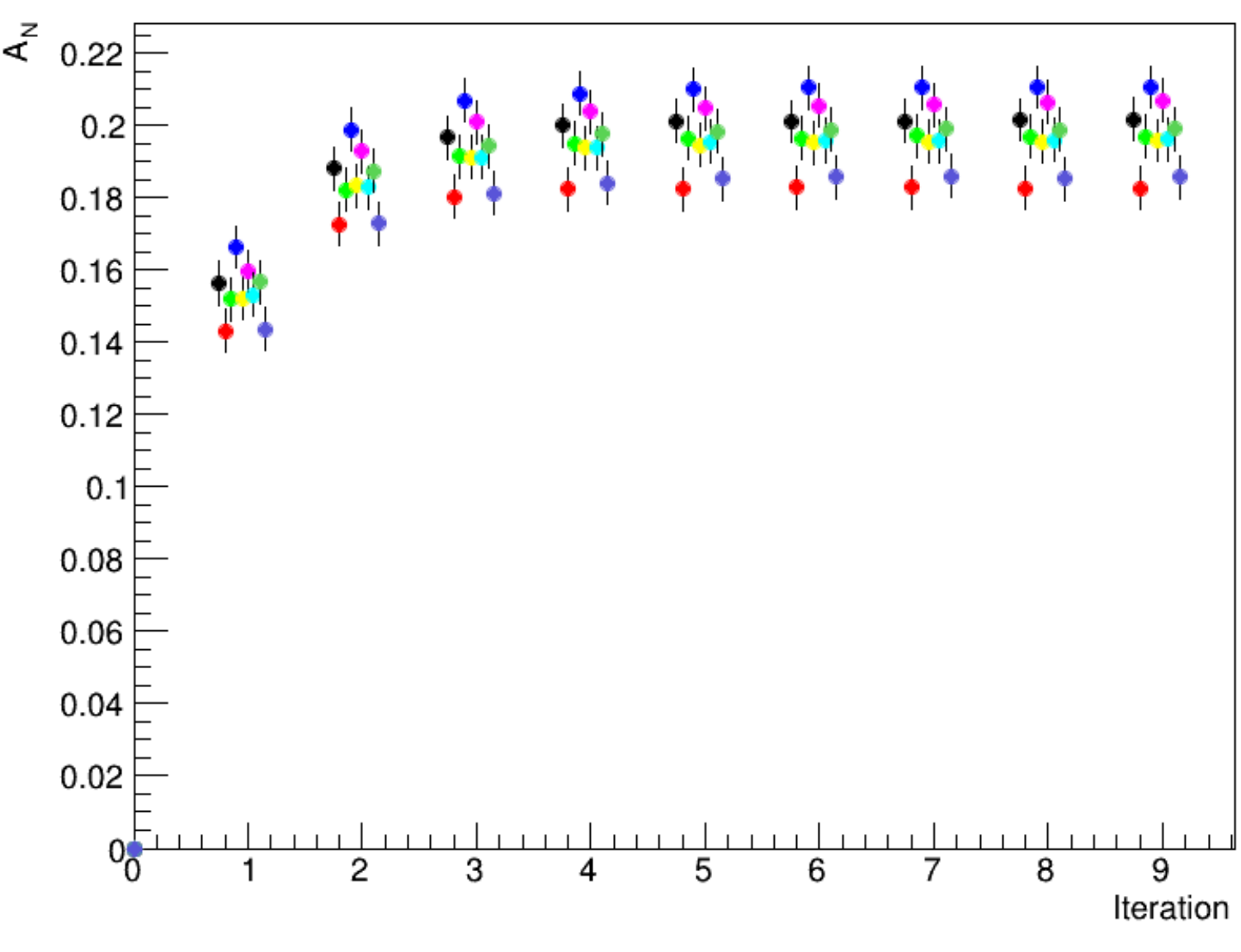}\\
Convergence test with strong smearing\\
\includegraphics[scale=0.4]{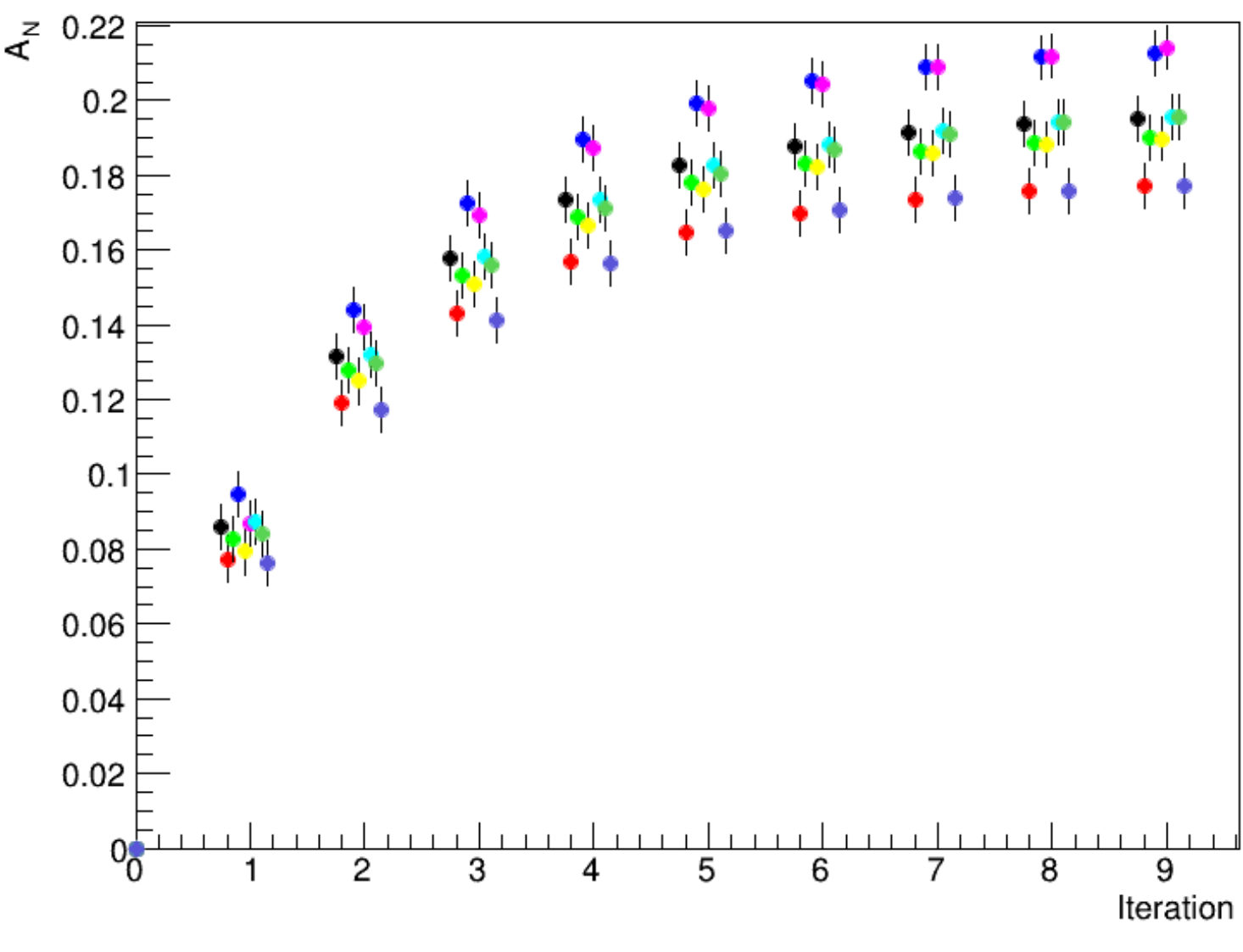}\\
\caption{Rates of convergence of the unbinned log-likelihood TSSA extraction.  The case of weak smearing is illustrated on the top, strong smearing on the bottom. Each color (black, red, blue, etc.) represents a statistically independent set of data used to test the unfolding; the points have been offset from each other slightly for visibiliy.  In all cases the embedded value of the asymmetry was $A_N=0.2$.  It is seen that the unfolding does not converge to the same answer in every dataset (black, red, blue, etc.); this effect is explored via the 250 tests described in Table~\ref{tab:250} and in Figures~\ref{fig:weak_histos} and \ref{fig:strong_histos}.}
\label{fig:iterations}
\end{figure}

In testing the reliability of the unfolding scheme, we were concerned with both the rate of convergence of unfolding and the possibility of a bias or systematic uncertainty in the results.  We tested both of these issues by observing the result of the TSSA extraction as a function of unfolding iteration number, for nine independent tests.  These results are displayed in Figure~\ref{fig:iterations}.  In these tests, the ``Simple'' set of parameters (see table~\ref{tab:setups}) was used to generate the events.  It is seen that in the case of weak smearing ($\sigma_{\rm smear}=0.45$ radians) the unfolding has converged in four iterations, while in the case of strong smearing ($\sigma_{\rm smear}=0.90$ radians) eight iterations are required.  In both cases, there is a systematic uncertainty on top of the statistical one, illustrated by the spread in results at each unfolding step. The systematic uncertainty is larger for the strong smearing case.

\begin{table}[ht]
\centering
\caption{Listing of results when each individual test was repeated 250 times.  The ``Mean $A_N$'' is the mean of the histogram of 250 test results for the foreground asymmetry, $\sigma_{250}$ is the width of that histogram, while $\sigma_1$ is the statistical uncertainty associated with a single result.  The Analysis Methods are: (1) Unfolding of Binned Data; (2) Binned Analysis of Unfolded Unbinned Data; and (3) Unbinned Log-Likelihood Analysis of Unfolded Unbinned Data.  These numerical data are extracted from the histograms in Figures~\ref{fig:weak_histos} and \ref{fig:strong_histos}.}
\label{tab:250}
\begin{tabular}{||L|L|L|L|L||}  \hline 
{\bf Tests with Weak Smearing} & Analysis Method & Mean $A_N$ & $\sigma_{250}$ & $\sigma_1$ \\
\hline \hline
\multirow{3}{4em}{``Simple''}
 & (1) &0.2016&0.0050&0.0044 \\
 & (2) &0.1935&0.0054&0.0043\\
 & (3) &0.1958&0.0053&0.0043\\ \hline
\multirow{3}{4em}{``Background Asymmetry''}
 & (1) &0.2014&0.0051&0.0044\\
 & (2) &0.1926&0.0056&0.0043\\
 & (3) &0.1947&0.0057&0.0043\\ \hline
\multirow{3}{4em}{``Pol/lumi Imbalance''}
 & (1) &0.2020&0.0068&0.0060\\
 & (2) &0.1933&0.0077&0.0059\\
 & (3) &0.1952&0.0077&0.0059\\ \hline
\multirow{3}{4em}{``Cosine Efficiency''}
 & (1) &0.2037&0.0141&0.0038\\
 & (2) &0.1950&0.0141&0.0059\\
 & (3) &0.1966&0.0138&0.0059\\ \hline \hline
{\bf Tests with Strong Smearing} & Analysis Method & Mean $A_N$ & $\sigma_{250}$ & $\sigma_1$ \\
\hline \hline
\multirow{3}{4em}{``Simple''}
 & (1) &0.1992&0.0066&0.0044\\
 & (2) &0.1906&0.0078&0.0044\\
 & (3) &0.1927&0.0079&0.0043\\ \hline
\multirow{3}{4em}{``Background Asymmetry''}
 & (1) &0.1990&0.0073&0.0044\\
 & (2) &0.1881&0.0085&0.0044\\
 & (3) &0.1902&0.0086&0.0044\\ \hline
\multirow{3}{4em}{``Pol/lumi Imbalance''}
 & (1) &0.1998&0.0091&0.0060\\
 & (2) &0.1895&0.0117&0.0059\\
 & (3) &0.1914&0.0116&0.0060\\ \hline
\multirow{3}{4em}{``Cosine Efficiency''}
 & (1) &0.1984&0.0166&0.0041\\
 & (2) &0.1889&0.0179&0.0059\\
 & (3) &0.1906&0.0170&0.0060\\ \hline
\end{tabular}
\end{table}

\begin{figure}[ht]
\centering
\includegraphics[scale=0.4]{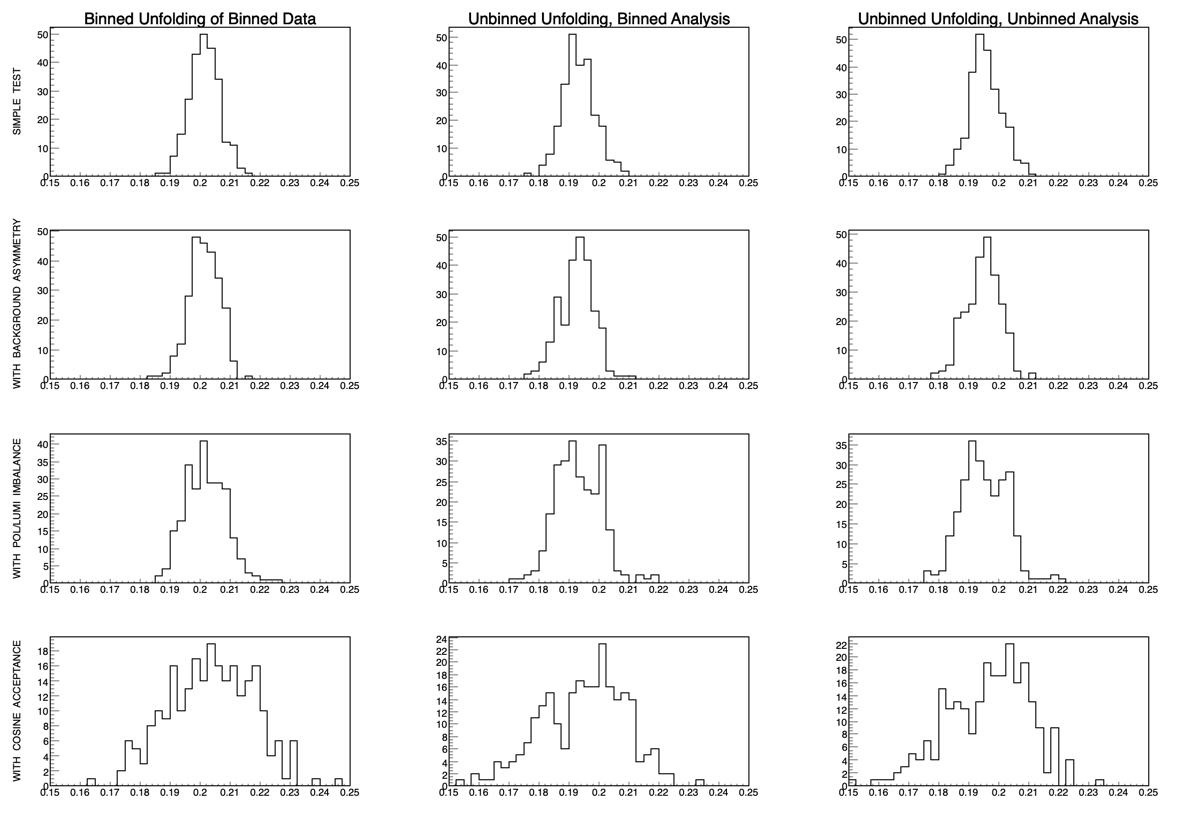}
    \caption{Tests with Weak Smearing:  Histograms of the extracted signal asymmetry resulting from 250 tests of the
    unfolding and analysis methods, when weak smearing was used on the simulated data.  Left to right are the cases of binned unfolding of binned data (analysis method 1), unbinned unfolding followed by binned analysis (method 2), and unbinned unfolding followed by an unbinned log-likelihood analysis (method 3).  Top to bottom are the ``simple'' test, then a test including a background asymmetry, then a test including also a polarization and luminosity imbalance, and finally a test including a cosine acceptance.  The conditions of these tests are detailed in Table~\ref{tab:background}. In every test the signal asymmetry was 0.20.  It is seen that as the tests become more difficult, the unfolding produces additional noise (the peak becomes wider) and bias (the peak is not centered on 0.20).  Numerical information about these histograms is listed in Table~\ref{tab:250}. }
    \label{fig:weak_histos}
\end{figure}

\begin{figure}[ht]
\centering
\includegraphics[scale=0.4]{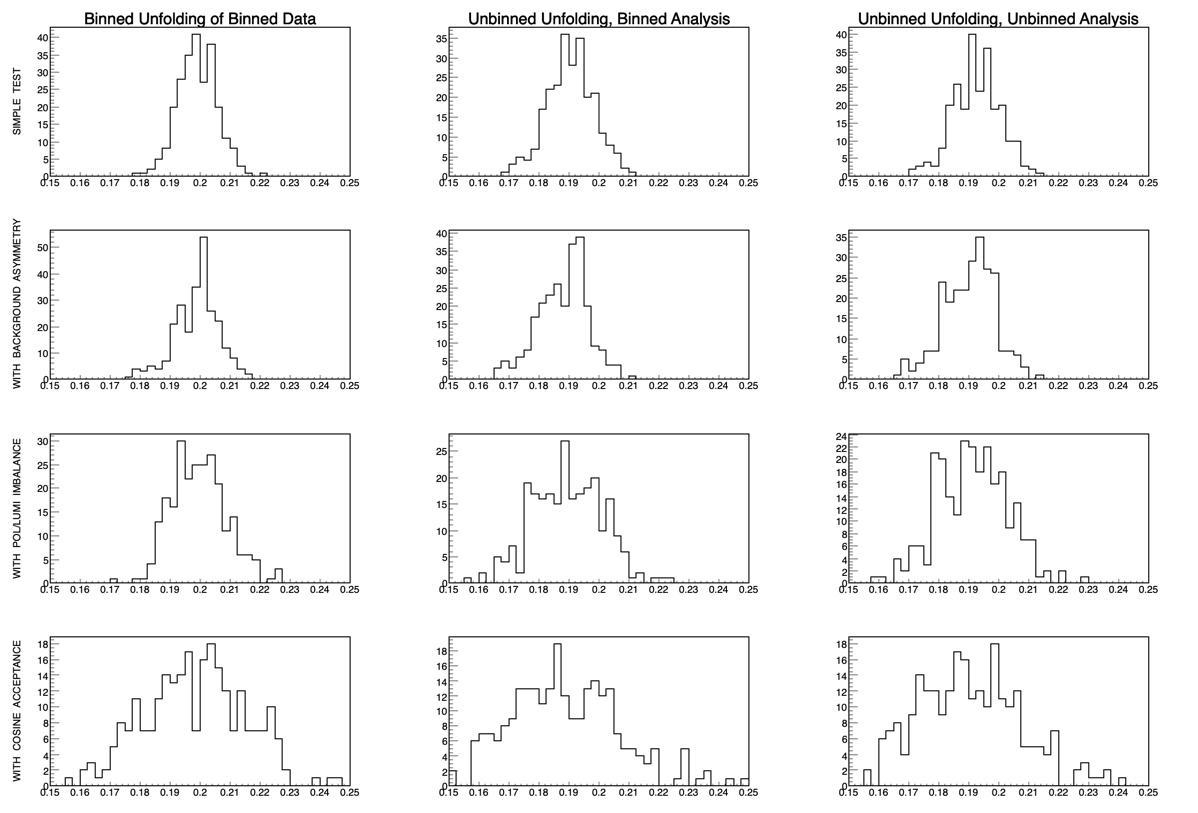}
    \caption{Tests with Strong Smearing:  Histograms of the extracted signal asymmetry resulting from 250 tests of the
    unfolding and analysis methods, when strong smearing was used on the simulated data.  Left to right are the cases of binned unfolding of binned data (analysis method 1), unbinned unfolding followed by binned analysis (method 2), and unbinned unfolding followed by an unbinned log-likelihood analysis (method 3).  Top to bottom are the ``simple'' test, then a test including a background asymmetry, then a test including also a polarization and luminosity imbalance, and finally a test including a cosine acceptance.  The conditions of these tests are detailed in Table~\ref{tab:background}. In every test the signal asymmetry was 0.20.  It is seen that as the tests become more difficult, the unfolding produces additional noise (the peak becomes wider) and bias (the peak is not centered on 0.20).  Numerical information about these histograms is listed in Table~\ref{tab:250}. }
    \label{fig:strong_histos}
\end{figure}

To quantify the systematic uncertainty associated with the unfolding, we performed 250 independent tests of both the weak and
strong smearing type under the four sets of conditions listed in table~\ref{tab:setups}.  For strong smearing we used 8 unfolding iterations, and for weak smearing we used 4 unfolding iterations. The data were analyzed in the three different methods mentioned earlier; unfolding of binned data, unfolding of unbinned data followed by a binned analysis, and log-likelihood analysis of unfolded unbinned data.  In each case, we histogrammed the 250 test results (see Figures~\ref{fig:weak_histos} and \ref{fig:strong_histos}) and compared the width of that histogram to the statistical uncertainty reported for an individual result.  The results of these tests are tabulated in table~\ref{tab:250}.  

In the tests with weak smearing, it is seen that all three analysis methods perform relatively the same.  All three experience some additional systematic error (reflected in the value of $\sigma_{250}$ as compared to $\sigma_1$) as the test data becomes more complicated, starting with the ``Simple'' test and evolving to the ``Cosine Efficiency'' test.  

On the other hand, in the tests with strong smearing, the unbinned unfolding has a greater systematic error as the tests become more complex; there is both a bias in the determined asymmetry (it is reported to be lower than the embedded value of 0.2) and there is increasing noise in the result.  The binned unfolding seems to perform better in these more difficult tests.

\section{Conclusions}

We have developed a completely general approach to the determination of a transverse single-spin asymmetry, including the possibility of unequal spin-up and spin-down polarization magnitudes and/or luminosities, and the possibility of a background that has its own asymmetry.   We have shown that both binned and unbinned methods can be used.  We have demonstrated its effectiveness with and without unfolding.  

We have shown that when the data is sufficiently complicated (due to polarization and luminosity imbalance, and difficult acceptance shapes) then the unfolding struggles to return an unbiased response.  Any analysis using these techniques will of course need to examine these kinds of effects.

Another kind of background that we did not consider here is a combinatoric background, such as one would encounter in an experiment observing di-leptons ($e^+e^-$ or $\mu^+\mu^-$ for example). Given a method for estimating the yield from this background (see for example \cite{SeaQuest:2023tcr}), then it could be accounted for using the same techniques described here.  In a binned method, it would simply be subtracted; in an unbinned method it would have a negative weight in a log-likelihood treatment.

Finally we make the obvious point that more than one transverse single-spin asymmetry may exist in the same scattering experiment; each one would have a different azimuthal-angle dependence: $\sin\phi$, $\cos\phi$, $\cos(2\phi)$, etc.  These can be simultaneously extracted using an extension of the techniques described here.  Of course in case of extraction of multiple parameters, then there will be covariances among them, and these can be determined using already well-established techniques.

\section{Acknowledgments}

This work was supported by grant DE-FG02-94ER40847 from the US Department of Energy, Office of Science, Experimental Medium Energy Nuclear Physics Program.  We are grateful to R.~Milton for advice in the use of the software at his GitHub website.

\clearpage

\bibliographystyle{model1-num-names.bst}

\bibliography{main}{}

\end{document}